\documentclass[epj]{svjour}
\usepackage{amsfonts}
\usepackage{graphicx}
\usepackage{amsmath}
\usepackage{amssymb}
\usepackage{subfigure}

\begin{document}

\title{The roles of quantum correlations in quantum cloning}
\author{Jun Zhang \and Shao-xiong Wu \and Chang-shui Yu\thanks{%
quaninformation@sina.com; ycs@dlut.edu.cn} }
\institute{School of Physics and Optoelectronic Technology, Dalian University of
Technology, Dalian 116024, China}
\date{Received: date / Revised version: date}

\abstract{
In this paper, we study the entanglement and quantum discord of the output modes in the
unified $1\rightarrow 2$ state-dependent cloning and probabilistic quantum
cloning. The tripartite entanglement among the output modes and the quantum cloning machine is also considered. We find that the roles of the quantum correlations including the bipartite and tripartite entanglement and quantum discord
strongly depend on the quantum cloning machines as well as the cloned state.
In particular, it is found that this quantum cloning scheme can be realizable
even without any quantum correlation.
 } 
\maketitle
\section{Introduction}

In recent years, quantum information theory has been developed greatly,
whilst a lot of quantum information processing tasks have been invented and
realized such as quantum teleportation \cite{tel}, quantum entanglement
purification \cite{purification1,purification2}, quantum key distribution
\cite{key,miyao} and so on \cite{cloningzong,cloningzongf,cloningzongjia}.
In particular, quantum key distribution, compared with the classical
cryptography, has the distinguished security which is guaranteed by the
fundamental principle, i.e., 'Quantum No-Cloning Theorem' (QNCT) \cite%
{nocloning}.

QNCT states that an unknown quantum state cannot be cloned (perfectly
copied) \cite{nocloning}, however, the approximate cloning can always be
realized based on some proper quantum schemes (quantum cloning machines).
The approximate cloning usually includes two types: one is the probabilistic
quantum cloning by which a set of linearly independent states can be
faithfully copied with a certain success probability \cite{PQC}, and the
other is the unfaithful quantum cloning by which the output state is not a
faithful copy of the input state \cite{SQC0,SQC}. Of course, the fidelity of
the output state in the unfaithful quantum cloning could depends on the
input state. The approximate quantum cloning has attracted widely interests
and there appears many interesting quantum cloning machines such as the
phase-covariant quantum cloning \cite%
{phase,phase1,phase2,phase3,phase4,phase5}, the optimal phase-independent
cloning \cite{jia4}, entanglement of quantum cloning \cite%
{jiu,jiu1,jiu2,jiu3}, the orthogonal qubit quantum cloning \cite{orth}, the
optimal nonorthogonal quantum cloning \cite{unified} and so on \cite%
{cloningzong,cloningzongf}. Among the various quantum clonings, Wang \textit{%
et al.} have proposed a unified way to construct general asymmetric
universal quantum cloning machine \cite{Ynwang}. In particular, the
Buzek-Hillery quantum cloning machine (BHC) has shown to be a universal one
because it achieves equal fidelity $\frac{5}{6}$ for any quantum state \cite%
{UQC}. However, in the BHC including many unfaithful cloning schemes, the
two output modes are usually correlated with each other. Compared with the
separable output modes in the ideal quantum cloning, intuitionally it seems
that the output correlated modes could play the negative role in the
cloning. That is, the success probability or the fidelity of the output
state could be reduced by the existing correlation. Such an intuition is
also supported by a recent job which shows, as a general viewpoint for the
imperfect cloning, that the amount of correlation between the original and
the cloned copy in the output will prevent from copying the information of
the original state in the blank state \cite{QCcloning}. On the contrary,
some other results show that in the probabilistic pure state cloning there
is no entanglement, but quantum correlation is always present \cite{support}%
. And for a certain class of optimal cloners, in order to reach higher fidelities, entanglement is needed \cite{support1}. So what are the roles of quantum correlations in the quantum cloning?

In this paper, we attempt to show that the roles of quantum correlations including the
concurrence, 3-tangle and quantum discord depend on the quantum
cloning scheme and the cloned states. Or from a general point of view, there does not exist one-to-one map between quantum cloning and quantum
correlation as well as quantum entanglement. In order to support our points,
we consider the optimal nonorthogonal quantum cloning machine which unifies
the $1\rightarrow 2$ state-dependent cloning and probabilistic quantum
cloning \cite{unified}. In this scheme, we find that given two states to be
cloned and an expected success probability (fidelity), the same fidelity
(success probability) could correspond to different quantum correlations. On
the contrary, a fixed quantum correlation could also corresponds to
different fidelities or success probabilities. Thus the most distinct
advantage is that for the fixed fidelity or success probability, one will
has diverse ways to control the cloning procedure with various quantum
correlations taken into account, or one will always find a way to implement
the cloning with the less cost, since the quantum correlations could be some
important physical resources. Finally, we also find that if there is no
quantum correlation in the scheme, the quantum cloning can also be realized.

This paper is organized as follows. In Sec. II, we give a brief introduction
of the unified $1\rightarrow 2$ quantum cloning. In Sec. III, we study how the success probability and the fidelity are influenced by the
bipartite and tripartite entanglement as well as  quantum discord. In Sec. IV, we
present a successful cloning without any quantum correlation. Finally, the
conclusion is drawn.

\section{Unified $1\rightarrow 2$ quantum cloning}

To begin with, let's give a brief introduction of the transformations in the
unified $1\rightarrow 2$ state-dependent cloning and probabilistic quantum
cloning. The action of this quantum cloning is given by the transformations
\cite{unified}%
\begin{eqnarray}
\left\vert 0\right\rangle \left\vert 0\right\rangle \left\vert
0\right\rangle _{p} &\rightarrow &\left[ A\left\vert 00\right\rangle
+B\left( \left\vert 01\right\rangle +\left\vert 10\right\rangle \right)
+C\left\vert 11\right\rangle \right] \left\vert 0\right\rangle _{p}  \notag
\\
&+&D\left\vert 00\right\rangle \left\vert 1\right\rangle _{p},  \notag \\
\left\vert 1\right\rangle \left\vert 0\right\rangle \left\vert
0\right\rangle _{p} &\rightarrow &\left[ A\left\vert 11\right\rangle
+B\left( \left\vert 01\right\rangle +\left\vert 10\right\rangle \right)
+C\left\vert 00\right\rangle \right] \left\vert 0\right\rangle _{p}  \notag
\\
&+&D\left\vert 00\right\rangle \left\vert 1\right\rangle _{p},
\label{transformation}
\end{eqnarray}%
where $A$, $B$, $C$, $D$ are real constants for simplicity and satisfy the
orthonormal condition:
\begin{eqnarray}
A^{2}+2B^{2}+C^{2}+D^{2} &=&1,  \label{tiao1} \\
2AC+2B^{2}+D^{2} &=&0.  \label{tiao2}
\end{eqnarray}%
with the subscript 'p' marking the probing state. Now, suppose we clone a
set of two nonorthogonal quantum states in the following form%
\begin{eqnarray}
\left\vert \chi _{1}\right\rangle &=&\cos \theta \left\vert 0\right\rangle
+\sin \theta \left\vert 1\right\rangle ,  \label{chu1} \\
\left\vert \chi _{2}\right\rangle &=&\sin \theta \left\vert 0\right\rangle
+\cos \theta \left\vert 1\right\rangle .  \label{chu2}
\end{eqnarray}%
where $\theta \in \lbrack 0,\pi /4],$ with their overlap given by%
\begin{equation}
s=\left\langle \chi _{1}|\chi _{2}\right\rangle =\sin 2\theta .
\label{overlap}
\end{equation}%
After the cloning transformations, the output states corresponding to the
two nonorthogonal quantum states are given, respectively, by
\begin{eqnarray}
\left\vert \chi _{1}\right\rangle ^{(out)} &=&[(A\cos \theta +C\sin \theta
)\left\vert 00\right\rangle  \notag \\
&&+B(\cos \theta +\sin \theta )(\left\vert 01\right\rangle +\left\vert
10\right\rangle  \notag \\
&&+(C\cos \theta +A\sin \theta )\left\vert 11\right\rangle ]\left\vert
0\right\rangle _{p}  \notag \\
&&+D(\cos \theta +\sin \theta )\left\vert 00\right\rangle \left\vert
1\right\rangle _{p}  \notag \\
&=&\sqrt{\gamma }\left\vert X_{1}\right\rangle \left\vert 0\right\rangle
_{p}+\sqrt{1-\gamma }\left\vert 00\right\rangle \left\vert 1\right\rangle
_{p},  \label{jie1} \\
\left\vert \chi _{2}\right\rangle ^{(out)} &=&[(A\sin \theta +C\cos \theta
)\left\vert 00\right\rangle  \notag \\
&&+B(\sin \theta +\cos \theta )(\left\vert 01\right\rangle +\left\vert
10\right\rangle  \notag \\
&&+(C\sin \theta +A\cos \theta )\left\vert 11\right\rangle ]\left\vert
0\right\rangle _{p}  \notag \\
&&+D(\sin \theta +\cos \theta )\left\vert 00\right\rangle \left\vert
1\right\rangle _{p}  \notag \\
&=&\sqrt{\gamma }\left\vert X_{2}\right\rangle \left\vert 0\right\rangle
_{p}+\sqrt{1-\gamma }\left\vert 00\right\rangle \left\vert 1\right\rangle
_{p},  \label{jie2}
\end{eqnarray}%
where $\sqrt{\gamma }\left\vert X_{i}\right\rangle =\left\langle
0_{p}\right\vert \left. \chi _{i}\right\rangle ^{(out)}$, $\gamma $ is the
normalization of $\left\langle 0_{p}\right\vert \left. \chi
_{i}\right\rangle ^{(out)}$ denoting the probability with which one will
obtains the $\left\langle 0_{p}\right\vert $ if one measures the probing
qubit. Note that $\gamma $ is a function of $A$, $B$, $C$, $D$ and $\theta $%
. If the probing state $\left\vert 0\right\rangle _{p}$ is detected by some
measurement, $\left\vert \chi _{i}\right\rangle ^{(out)}$ will collapses to
the state $\left\vert X_{i}\right\rangle $ with the probability $\gamma $,
which means that this cloning is not only probabilistic but also
state-dependent and the fidelity is given by the overlap between the output
state $\mathrm{Tr}_{\alpha }\left\{ \left\vert X_{i}\right\rangle
\left\langle X_{i}\right\vert \right\} $ and the input state $\left\vert
\chi _{i}\right\rangle $, i.e.,
\begin{eqnarray}
f &=&\frac{1}{4\gamma }[(A-2B-C)(A+C)\cos 4\theta +2BC+C^{2}+3A^{2}  \notag
\\
&&+4(A+B)(B+C)\sin 2\theta +2AB+4B^{2}],  \label{fidelity}
\end{eqnarray}%
where $\alpha $ in the subscript denotes trace over either subsystem. A
simple calculation will shows that this cloning is symmetric because the
fidelity $f$ pertains to both $\left\vert X_{1}\right\rangle $ and $%
\left\vert X_{2}\right\rangle $. In addition, this fidelity $f$ is not an
optimal one. The optimal case will be discussed in the latter section. If $%
\left\vert 1\right\rangle _{p}$ is obtained for some measurement, it means
that the cloning fails. If the success probability $\gamma =1$, the cloning
is called the state-dependent cloning (SDC), otherwise, the cloning is
considered as a probabilistic one with the success probability denoted by $%
\gamma ^{(PQC)}=\gamma $.

\section{Quantum correlation in the quantum cloning}

\subsection{Quantum entanglement}

In order to find the relation between the quantum entanglement and the
fidelity, we have to first deal with the fidelity given by Eq. (\ref%
{fidelity}). Consider Eqs.(\ref{tiao1},\ref{tiao2},\ref{overlap},\ref{jie1},%
\ref{jie2}), we can find that $D=\sqrt{\frac{1-\gamma }{1+s}}$ and the
values of $A$ and $C$ are given by
\begin{equation}
\left\{
\begin{array}{c}
A_{1\pm}=\frac{1}{2}\left( \pm \sqrt{\frac{s+2\gamma -1}{1+s}-4B^{2}}%
+1\right) , \\
C_{1\pm}=\frac{1}{2}\left( \pm \sqrt{\frac{s+2\gamma -1}{1+s}-4B^{2}}%
-1\right) .%
\end{array}%
\right.  \label{x1}
\end{equation}%
or%
\begin{equation}
\left\{
\begin{array}{c}
A_{2\pm}=\frac{1}{2}\left( \mp \sqrt{\frac{s+2\gamma -1}{1+s}-4B^{2}}%
-1\right) , \\
C_{2\pm}=\frac{1}{2}\left( \mp \sqrt{\frac{s+2\gamma -1}{1+s}-4B^{2}}%
+1\right) .%
\end{array}%
\right.  \label{x2}
\end{equation}%
We would like to emphasize $\gamma \geq \frac{1-s}{2}$ and $B\in \left[ -%
\frac{\sqrt{1-2D^{2}}}{2},\frac{\sqrt{1-2D^{2}}}{2}\right] $ due to the real
$A$, $C$. This condition will be used throughout of this paper. Insert Eqs. (%
\ref{x1},\ref{x2}) into Eq. (\ref{fidelity}), through a long and tedious
procedure one will finds that the general (unoptimized) fidelity reads%
\begin{equation}
f_{1\pm}=\frac{1}{2}\pm \frac{(1+s)[1+(2B-1)s]}{2\gamma }\sqrt{\frac{%
s-1-4B^{2}(1+s)+2\gamma }{1+s}}.  \label{f1}
\end{equation}%
or%
\begin{equation}
f_{2\pm}=\frac{1}{2}\mp \frac{(1+s)[-1+(2B+1)s]}{2\gamma }\sqrt{\frac{%
s-1-4B^{2}(1+s)+2\gamma }{1+s}}.  \label{f2}
\end{equation}%
Since it is always implied that the maximal output fidelity is expected for any $s$, $\gamma$ and $B$, one can find that such a partially optimal fidelity
can be given by $f_{p}=\max\{f_{1\pm},f_{2\pm}\}$. Thus for a pair of fixed input states (fixed $s$) and the expected success probability $\gamma$, $f_{p}$ is a function of the parameter $B$, from which one can find out how to increase the fidelity $f_{p}$ by adjusting $B$. In addition, it will be shown that the parameter $B$ of the cloners is important for the
connection with entanglement.

In the following, we will find out the amount of entanglement hidden in the
two output modes. Due to the symmetry of the transformation given by Eq. (%
\ref{transformation}), we will only consider the state $\left\vert \chi
_{1}\right\rangle ^{out}$. It can be easily shown that $\left\vert \chi
_{2}\right\rangle ^{out}$ has the same result which is not repeated. The
state of the two output modes can be easily obtained from Eqs. (\ref{jie1},%
\ref{jie2}), by tracing over the probing qubit $p$. So the reduced density
matrix of the two output modes can be given in the following form:
\begin{equation}
\rho =\left(
\begin{array}{cccc}
a^{2}+d^{2} & ab & ab & ac \\
ab & b^{2} & b^{2} & bc \\
ab & b^{2} & b^{2} & bc \\
ac & bc & bc & c^{2}%
\end{array}%
\right) .  \label{density}
\end{equation}%
with%
\begin{eqnarray}
a &=&A\cos \theta +C\sin \theta , \\
b &=&B(\cos \theta +\sin \theta ), \\
c &=&C\cos \theta +A\sin \theta , \\
d &=&D(\cos \theta +\sin \theta ).
\end{eqnarray}%
Note that $A$, $B$, $C$, $D$ are determined by Eqs. (\ref{x1},\ref{x2}). In
order to quantify the entanglement in the state $\rho $, we would like to
employ the concurrence as the entanglement measure which is defined for a
density matrix $\rho $ as \cite{concurrence}
\begin{equation}
C\left( \rho \right) =\max \{0,\sqrt{\lambda _{1}}-\sqrt{\lambda _{2}}-\sqrt{%
\lambda _{3}}-\sqrt{\lambda _{4}}\},  \label{enta}
\end{equation}%
where $\lambda _{i}$ is the square root of the $i$th eigenvalue of the
matrix $\rho \sigma _{y}\otimes \sigma _{y}\rho ^{\ast }\sigma _{y}\otimes
\sigma _{y}$ in decreasing order with $\sigma _{y}$ denoting the Pauli
matrix. Substitute the density matrix $\rho $ into Eq. (\ref{enta}), one
will obtain that
\begin{equation}
\lambda _{1,2}=0,\lambda _{3,4}=\alpha \pm \beta .
\end{equation}%
with
\begin{eqnarray}
\alpha &=&2(b^{2}-ac)^{2}+c^{2}d^{2}, \\
\beta &=&2\left\vert b^{2}-ac\right\vert \sqrt{(b^{2}-ac)^{2}+c^{2}d^{2}}.
\end{eqnarray}%
So the concurrence of the output state will be given by
\begin{equation}
C(\rho )=\max \{\left\vert \sqrt{\lambda _{3}}-\sqrt{\lambda _{4}}%
\right\vert ,0\}.  \label{jiuchan}
\end{equation}%
It is obvious that the concurrence depends on the parameters $A$, $B$, $C$, $%
D$. Since $A$ and $C$ can be given as functions of $B$ due to Eqs. (\ref{x1},%
\ref{x2}), $C(\rho )$ also depends on $B$. Unfortunately, the relation
between $C(\rho )$ and $B$ is too complicated to give in an explicit form,
we will have to consider it in a numerical way.

In principle, one can grasp any relation between the fidelity and the
concurrence that depend on the common parameter $B$. In Fig.1, we plot how
the partially optimal fidelity $f_{p}$ in Eqs. (\ref{f1},\ref{f2}) varies with the concurrence
when the parameter $\theta $ of the input state is chosen as $0$, $\pi /20$,
$\pi /10$, $\pi /4$ for different $\gamma $. From this figure, one could say
that the large concurrence will lead to the decay of the fidelity for $%
\theta =0$, however, for the other values of $\theta $, there does not exist
a one-to-one map between concurrence and the fidelity. It is obvious that
the  fidelity $f_{p}$ could keep invariant, go up or go down for two
different concurrence. In particular, this not only depends on the overlap
between the two cloned states, but also depends on the probability. This can
be seen from Fig. 1. It is interesting that, with the increase of $\theta $,
one can find that the single curve (for $\theta =0$) will split into two
curves with the same left end point. In particular, when $\theta =\pi /4$, the two curves almost coincide (we have carefully calculate
the two curves that are too close to show explicitly in the figure). In
addition, for different $\gamma $, the figures are similar, but the most obvious
feature is that the small $\gamma $ will limit the maximal concurrence that
could be produced in the cloning. Therefore, for the general case, our
conclusion is that the entanglement can not uniquely determine of the
general fidelity.
\begin{figure}[tbp]
\includegraphics[width=1\columnwidth]{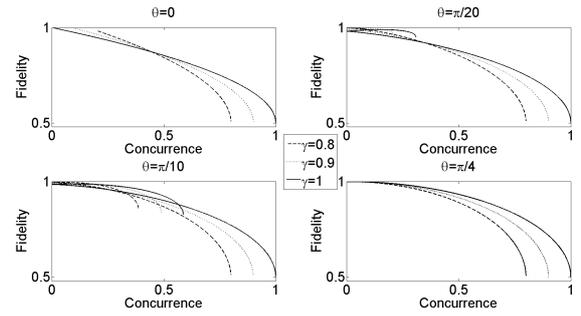}
\caption{The partially optimal fidelity $f_{p}$ versus the concurrence. $\protect\theta $
takes $0$,$\protect\pi /20$,$\protect\pi /10$,$\protect\pi /4$ for each
subplot, respectively, and $\protect\gamma =1,0.9,0.8$ for different line
styles.}
\end{figure}
\begin{figure}[tbp]
\includegraphics[width=1\columnwidth]{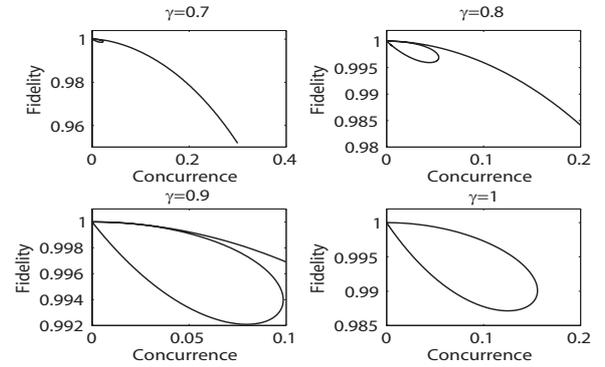}
\caption{The optimal fidelity $f_{(opt)}$ versus the concurrence with $%
\protect\gamma $ takes $0.7$, $0.8$, $0.9$, $1$, respectively.}
\end{figure}

In the following, we will discuss the relation between quantum entanglement
and the optimal fidelity for the unified $1\rightarrow 2$ quantum cloning.
From the partially optimal fidelity (\ref{f1}) and (\ref{f2}). Using the method of
Lagrange multipliers, it is easy to derive an explicit expression from $%
\frac{\partial f_{i}(B)}{\partial B}=0$. The reasonable value of parameter $%
B $ reads%
\begin{equation}
B_{1}=\frac{s^{2}-1+M}{8(s+s^{2})}.  \label{b1}
\end{equation}%
and%
\begin{equation}
B_{2}=-\frac{s^{2}-1+M}{8(s+s^{2})}.  \label{b2}
\end{equation}%
with $M=\sqrt{1+s^{2}[9s^{2}+16(1+s)\gamma -10]}$. Thus, replacing the $B$
with (\ref{b1}) and (\ref{b2}) in Eqs. (\ref{f1},\ref{f2}), respectively, the above two partially optimal
 fidelities will arrive at a common thoroughly optimal fidelity%
\begin{equation}
f_{(opt)}=\frac{1}{2}+\frac{(3+M-3s^{2})}{32\gamma }\sqrt{\frac{%
[2(s-1)(1-M+3s^{2})+16s^{2}\gamma ]}{s^{2}(1+s)}}.
\end{equation}%
In Fig 2, the plots show that how the optimal fidelity $f_{(opt)}$ varies with the
concurrence when the parameter $\gamma $ takes $0.7$, $0.8$, $0.9$, $1$. It
is clear that even though for some particular $\gamma $, the concurrence and
the fidelity show the converse trends, in general, they are not the
one-to-one correspondence between them. That is, a fixed fidelity or
probability could corresponds to different concurrence. In other words, a
fixed concurrence could be enough for some input states to realize the cloning with some expectations (such as the fidelities or
probabilities), but it could not enough for other states to realize the cloning with the same expectations.
Compared with Fig. 1, one will find that the optimal fidelity corresponds to the narrow range of concurrence (not more than 0.4).

\subsection{Quantum discord}

As an important quantum correlation beyond quantum entanglement, quantum
discord has been thought to be another physical resource in some quantum
information processing tasks \cite{zong} . Recently some results have proven
that in some quantum cloning, quantum discord is always present \cite%
{support}. Next, we will employ the original quantum discord as a quantum
correlation measure to study the quantum correlation in the unified $%
1\rightarrow 2$ quantum cloning. Quantum discord $Q(\rho _{AB})$ introduced
by Ollivier and Zurek \cite{Olliver}, Vedral \textit{et. al}. \cite{Vedral},
respectively, is defined based on the loss of the information induced by
some optimal measurements on only one subsystem%
\begin{equation}
Q(\rho _{AB})=\min_{\Pi _{i}^{B}}S_{\{\Pi _{i}^{B}\}}(\rho _{A|B})+S(\rho
_{B})-S(\rho _{AB}),\label{discord}
\end{equation}%
where $S(\cdot )$ represents the von Neumann entropy and
\begin{equation}
S_{\{\Pi _{i}^{B}\}}(\rho _{A|B})=\sum\nolimits_{j}q_{j}S(\rho _{A}^{j}),
\end{equation}%
with the post-measured state $\rho _{A}^{j}$ given by%
\begin{equation}
\rho _{A}^{j}=\frac{\mathrm{Tr}_{B}[(I_{A}\otimes \Pi _{j}^{B})\rho
_{AB}(I_{A}\otimes \Pi _{j}^{B})]}{\mathrm{Tr}_{AB}[(I_{A}\otimes \Pi
_{j}^{B})\rho _{AB}(I_{A}\otimes \Pi _{j}^{B})]},  \label{3}
\end{equation}%
and $q_{j}$ the corresponding probability given by%
\begin{equation}
q_{j}=\mathrm{Tr}_{AB}[(I_{A}\otimes \Pi _{j}^{B})\rho _{AB}(I_{A}\otimes
\Pi _{j}^{B})].
\end{equation}%
It is noted that $\{\Pi _{j}^{B}\}$ is a complete set of projective
measurements.

Substitute Eq. (\ref{density}) into Eq. (\ref{discord}), we can obtain the relation between the
fidelity and the quantum discord. However, it is unfortunate that quantum discord given in Eq. (\ref{discord})
has not an analytic form for a general quantum state. So we have to numerically calculate the discord. In Fig. 3, it shows that how
the partially optimal fidelity $f_{p}$ in Eqs. (\ref{f1},\ref{f2}) varies with the quantum discord and the input
state with the parameter $\theta $ chosen as $0$, $\pi /20$, $\pi /10$, $\pi
/4$ for $\gamma =0.8,0.9,1$. From this figure, we also learn that the
partially optimal fidelity and the quantum discord do not have a one-to-one
correspondence. This is much like Fig. 1. However, one can note that when $%
\theta =\pi /4$, the two curves for $\gamma =0.8,0.9$ are separated
clearly. While in Fig. 4, it shows
that how the optimal fidelity $f_{(opt)}$ varies with quantum discord when
the parameter $\gamma $ takes $0.7$, $0.8$, $0.9$, $1$. From these figures,
we can see that the quantum discord is not the unique factor that influences the fidelity of
the unified $1\rightarrow 2$ quantum cloning.
\begin{figure}[tbp]
\includegraphics[width=1\columnwidth]{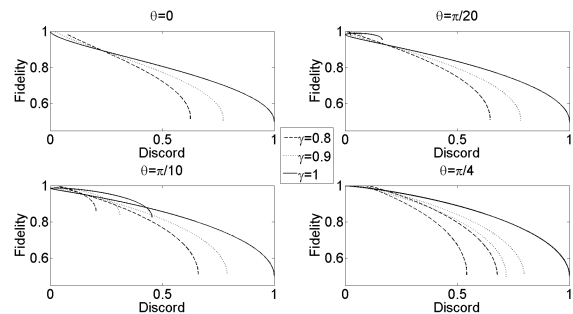}
\caption{The partially optimal fidelity $f_{p}$ versus the quantum discord. $\protect%
\theta $ takes $0$,$\protect\pi /20$,$\protect\pi /10$,$\protect\pi /4$ for
each subplot, respectively, and $\protect\gamma =1,0.9,0.8$ for different
line styles.}
\end{figure}
\begin{figure}[tbp]
\includegraphics[width=1\columnwidth]{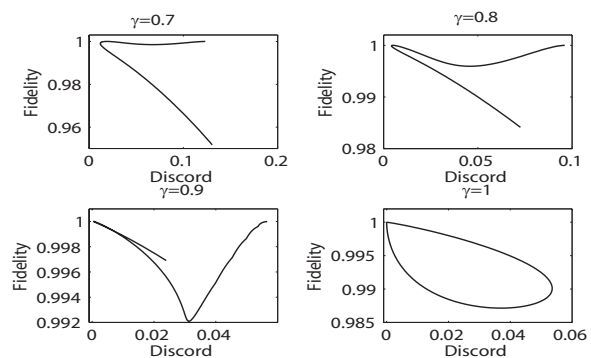}
\caption{The optimal fidelity $f_{(opt)}$ versus the quantum discord with $%
\protect\gamma $ takes $0.7$, $0.8$, $0.9$, $1$, respectively.}
\end{figure}

\subsection{3-tangle}

We've studied the bipartite correlation of the two output modes in the
quantum cloning machine. The result shows that quantum correlation mainly depends on the quantum cloning scheme
and the cloned states. However, the whole cloning procedure includes three
parties. So, we would like to study the role of tripartite entanglement
in the cloning machine. For a pure state $\left\vert \psi \right\rangle
_{ABC}$ of three qubits, the 3-tangle $\tau \left( \left\vert \psi
\right\rangle _{ABC}\right) $ is a good entanglement measure which can be
given by [35,36]
\begin{equation}
\tau \left( \left\vert \psi \right\rangle _{ABC}\right) =\sqrt{\left[ Tr\rho
_{AB}\tilde{\rho}_{AB}\right] ^{2}-Tr\left[ \rho _{AB}\tilde{\rho}_{AB}%
\right] ^{2}},  \label{tangle}
\end{equation}%
where $\tilde{\rho}_{AB}=\sigma _{y}\otimes \sigma _{y}\rho _{AB}^{\ast
}\sigma _{y}\otimes \sigma _{y}$ with $\rho _{AB}$ the reduced density
matrix by tracing over party $C$. Thus substitute Eqs. (10), (11) and (14) into Eq. (\ref{tangle}), one will find that the 3-tangle will arrives
at
\begin{eqnarray}
\tau (\rho _{ABC}) &=&\frac{(1-\gamma )}{\sqrt{2}}\left[ \gamma
-2B^{2}(1+\sin 2\theta )\right.  \notag \\
&&-\left. \cos 2\theta \sqrt{1-4B^{2}-\frac{2(1-\gamma )}{1+\sin 2\theta }}%
\right] .  \label{3an}
\end{eqnarray}%
In particular, one will easily check that Eq. (10) and Eq. (11) will lead to
the same result, i.e., Eq. (\ref{3an}). Thus for any $\gamma \geq \frac{1-s}{%
2}$, there always exists $B$ that can connect the 3-tangle with the partially optimal
fidelity $f_{p}$. Analogously, for the optimal cloning, the optimal fidelity
$f_{(opt)}$ can also be related with the 3-tangle by $s$.
\begin{figure}[tbp]
\includegraphics[width=1\columnwidth]{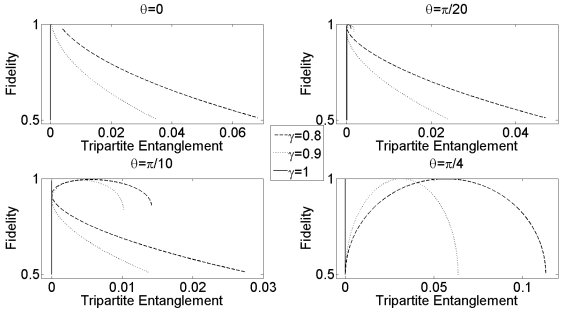}
\caption{The partially optimal fidelity $f_{p}$ versus the tripartite entanglement
(3-tangle $\protect\tau $). $\protect\theta $ takes $0$,$\protect\pi /20$,$%
\protect\pi /10$,$\protect\pi /4$ for each subplot, respectively, and $%
\protect\gamma =1,0.9,0.8$ for different line styles.}
\end{figure}
\begin{figure}[tbp]
\includegraphics[width=1\columnwidth]{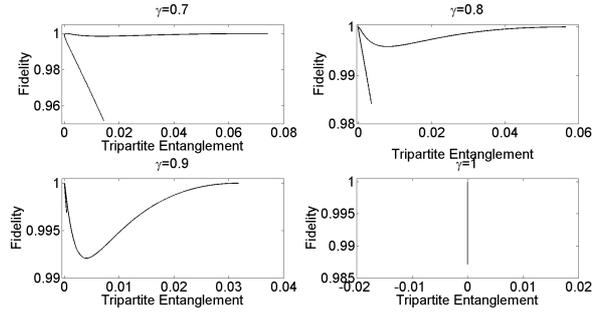}
\caption{The optimal fidelity $f_{(opt)}$ versus the tripartite entanglement
(3-tangle $\protect\tau $) with $\protect\gamma $ takes $0.7$, $0.8$, $0.9$,
$1$, respectively}
\end{figure}
We have illustrated the relations between the partially optimal fidelity $f_{p}$ as
well as the optimal fidelity $f_{(opt)}$ and the 3-tangle in Fig. 5. and
Fig. 6. From the two figures, it is found that the 3-tangle vanishes when $%
\gamma =1$ (the vertical line in the figures), which can be easily
understood since $\gamma =1$ corresponds to a product output state. One can
easily find that these relations are analogous to those for bipartite
entanglement. That is,  the same 3-tangle could corresponds to different
fidelities or probabilities.

\section{Without any quantum correlation}

From the above discussions, one can see that there is not a one-to-one
relation between the fidelity and the various quantum correlations. The
remaining question is whether quantum correlation is necessary for a
successful quantum cloning. It is interesting that, in the unified $%
1\rightarrow 2$ quantum cloning, quantum discord is actually not necessary
at all. Let us choose some special constraint conditions on the reduced
density matrix in the Eq. (\ref{density}) as follows.
\begin{eqnarray}
\gamma &=&1 \\
b^{2} &=&ac.
\end{eqnarray}%
Thus, the reasonable values of parameters $A$ and $C$ can be given by
\begin{equation}
\left\{
\begin{array}{c}
A_{1}^{\prime }=\frac{1+s+\sqrt{(1+s)}}{2+2s}, \\
B_{1}^{\prime }=\frac{\sqrt{s}}{2\sqrt{1+s}}, \\
C_{1}^{\prime }=\frac{\sqrt{(1+s)}-(1+s)}{2+2s}.%
\end{array}%
\right.  \label{no1}
\end{equation}%
\begin{equation}
\left\{
\begin{array}{c}
A_{2}^{\prime }=-\frac{1+s+\sqrt{(1+s)}}{2+2s}, \\
B_{2}^{\prime }=-\frac{\sqrt{s}}{2\sqrt{1+s}}, \\
C_{2}^{\prime }=\frac{(1+s)-\sqrt{(1+s)}}{2+2s}.%
\end{array}%
\right.  \label{no2}
\end{equation}%
Thus, the state of the two output modes will be reduced to
\begin{equation}
\mathrm{Tr}_{p}\left\vert \chi _{1}\right\rangle ^{(out)}\left\langle \chi
_{1}\right\vert =\left\vert \widetilde{\chi _{1}}\right\rangle \left\langle
\widetilde{\chi _{1}}\right\vert \otimes \left\vert \widetilde{\chi _{1}}%
\right\rangle \left\langle \widetilde{\chi _{1}}\right\vert .
\end{equation}%
with $\left\vert \widetilde{\chi _{1}}\right\rangle =\left( \sqrt{a}%
\left\vert 0\right\rangle \pm \sqrt{c}\left\vert 1\right\rangle \right) .$
It is obvious that the two output modes do not have either the quantum
entanglement or quantum discord. Insert the Eqs. (\ref{no1}, \ref{no2}) into
Eq. (\ref{fidelity}), we will find that the cloning fidelity reads
\begin{equation}
f_{(no)}=\frac{1}{2}\left[ 1+s^{\frac{3}{2}}+(1-s)\sqrt{1+s}\right] .
\end{equation}%
Although there is not any quantum correlation in the unified $1\rightarrow 2$
quantum cloning, the quantum cloning machine that determined by the
parameters Eqs. (\ref{no1}, \ref{no2}) will still arrives at a large
fidelity. One is able to find that the minimal value will achieve $0.9811$
when the $s$ is $0.3333$. Finally, we would like to emphasize that even
though there is no quantum correlation in the cloning procedure, it is
different from the classical cloning. The reason is that in the classical
world, the overlap $s$ for the two cloned states $\left\vert \chi
_{1}\right\rangle $ and $\left\vert \chi _{2}\right\rangle $ must be $s=0,1$
which corresponds to the unit fidelity $f_{(no)}$, but it is not the case in
the quantum world.

\section{Discussions and Conclusion}

We have studied the various quantum correlations including the concurrence,
quantum discord and 3-tangle in the unified $1\rightarrow 2$ state-dependent
cloning and probabilistic quantum cloning. The results show that these
quantum correlations in quantum cloning can not uniquely determine the
characteristic parameters of the cloning such as the fidelity and success
probability. These parameters depend on not only the various quantum
correlations but also the quantum cloning scheme and the cloned states. In
other words, for a fixed success probability or the fidelity, one could
implement the quantum cloning with different quantum correlations. This
could provide us a way to realize the quantum cloning with low cost
(quantum correlations). In particular, we even find  that
the successful quantum cloning does not need any quantum correlation at all.

Finally, we would like to emphasize that all that we have studied are
restricted to the case with real parameters $A,B,C,D$. It is natural that
the case with the complex numbers will become quite complicated. However, we
conjecture that the situation with complex parameters will lead to the
analogous conclusion as our current results.

\section{Acknowledgement}

This work was supported by the National Natural Science Foundation of China,
under Grants No.11375036 and 11175033.

%
\bibliographystyle{Style/icpig}
\bibliography{MicroPhys}

\end{document}